\documentclass[aps,prl,twocolumn,showpacs]{revtex4}
\usepackage{graphicx} 
\usepackage{dcolumn}  
\usepackage{bm}       

\begin{document}
\title{Full transmission through perfect-conductor subwavelength hole arrays}
\author{F.~J.~Garc\'{\i}a~de~Abajo,$^{1,2}$ R.
G\'{o}mez-Medina,$^{2,3}$ and J.~J. S\'{a}enz$^3$}
\address{$^1$Unidad de F\'{\i}sica de Materiales CSIC-UPV/EHU
Aptdo. 1072, 20080 San Sebastian, Spain \\
$^2$Donostia International Physics Center (DIPC), Paseo Manuel de
Lardiz\'{a}bal, 4, 20018 San Sebastian, Spain \\
$^3$Deptartamento de F\'{\i}sica de la Materia Condensada,
Universidad Aut\'{o}noma de Madrid, Cantoblanco, 28049 Madrid,
Spain}

\date{\today}

\begin{abstract}
Light transmission through 2D subwavelength hole arrays in
perfect-conductor films is shown to be complete (100\%) at some
resonant wavelengths even for arbitrarily narrow holes.
Conversely, the reflection on a 2D planar array of non-absorbing
scatterers is shown to be complete at some wavelengths regardless
how weak the scatterers are. These results are proven analytically
and corroborated by rigorous numerical solution of Maxwell's
equations. This work supports the central role played by dynamical
diffraction during light transmission through subwavelength hole
arrays and it provides a systematics to analyze more complex
geometries and many of the features observed in connection with
transmission through hole arrays.
\end{abstract}



\maketitle


The fraction of light transmitted through a periodic array of
subwavelength holes perforated in a metallic film has been shown
to exceed the open fraction occupied by the holes at certain
wavelengths related to the array periodicity \cite{ELG98}, and the
observed transmission can be anomalously large as compared to well
established predictions for isolated holes \cite{B1944}. This
phenomenon has triggered an intense amount of activity on both
experimental \cite{ELG98,GSH03,KES04,GBM04,BMD04,MH04,CN04,LT04}
and theoretical \cite{T99,PNE00,MGL01,W01,SGZ01,SVV03,KES04,LT04}
fronts. While the first observations were made in the near
infrared (NIR) \cite{ELG98,KES04,GBM04,BMD04,LT04}, the effect has
been recently corroborated at THz frequencies
\cite{GSH03,MH04,CN04}, where claims have been made that the
transmission is even larger \cite{MH04,CN04}. Two different kinds
of complementary interpretations of the enhanced transmission have
been proposed so far, depending on whether surface plasmons
\cite{ELG98,PNE00,MGL01,W01,SGZ01,BMD04} or dynamical diffraction
resonances \cite{T99,SVV03,LT04} are invoked as the origin of the
effect.

\begin{figure}
\centerline{\scalebox{0.45}{\includegraphics{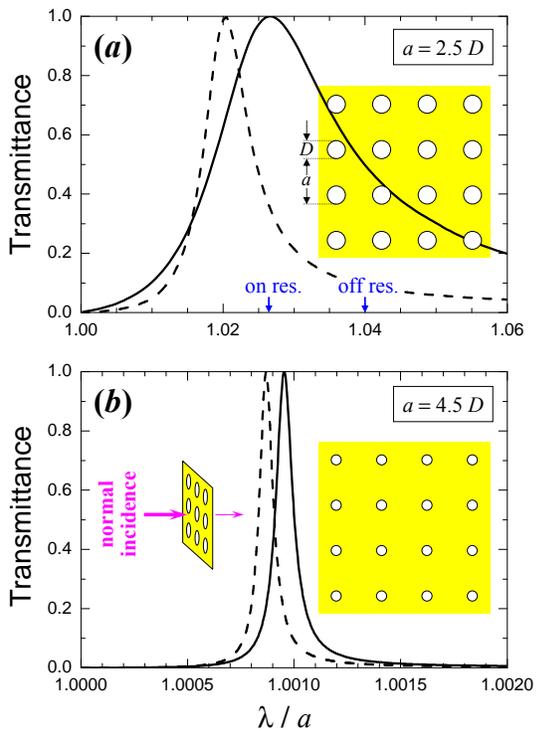}}}
\caption{\label{Fig1} Light transmittance through a square array of
circular holes perforated in a thin perfect-conductor screen as a
function of wavelength $\lambda$ normalized to the lattice constant
$a$ for two different diameters of the holes: {\bf (a)} $D=a/2.5$
and {\bf (b)} $D=a/4.5$. The light is impinging normal to the screen
and a 100\% transmission peak is obtained in both cases. Solid
curves: full numerical results. Broken curves: analytical model of
Eq.\ (\ref{dipoles}).}
\end{figure}

In this Letter, Babinet's principle \cite{J1975} is used to study
light transmission through hole arrays in perfect-conductor thin
films (PCTFs) by relating it to the reflection on planar arrays of
metallic disks, which are solved from the multipolar
polarizability of single disks. It is found that transmission
reaches 100\% at some resonant wavelengths regardless how small
the holes are, while the complementary system exhibits perfect
reflection resonances for arbitrarily small disks. These
resonances are associated to divergences in the coherent
interaction among disks/holes, as in the Wood anomalies
\cite{W1935}. A quasi-bound-state (QBS) close to the onset of the
first propagating diffraction channel (i.e., close to a Rayleigh
frequency) shows up due to the interaction of the scattered field
with diffraction modes. The coupling between the incoming field
and the QBS leads to transmission/reflection resonances which
display characteristic Fano-line shapes \cite{F1961}. This is
rigorously proven by analytically solving the small-hole limit and
also via full numerical solution of Maxwell's equations.
A typical result of our analysis is advanced in Fig.\ \ref{Fig1},
showing full transmission at wavelengths $\lambda$ immediately
above the lattice period $a$. The theory of hole arrays in PCTFs
traces back to the works of Eggimann, Collin, and Chen
\cite{Eggimann}, and McPhedran {\it et al.} \cite{McPhedran}. The
latter presents comprehensive comparisons with experiments,
demonstrating the existence of 100\% transmission peaks at
wavelengths immediately above the period. Here, we extend the
result of full transmission to arbitrarily small holes.

Babinet's principle \cite{J1975} provides a powerful tool for
analyzing complex perfect-conductor planar structures. For
instance, it implies that any self-complementary 2D-patterned flat
screen (e.g., a chess-board of square apertures) reflects and
transmits exactly half of the unpolarized light coming at normal
incidence for any wavelength. Also, a simple derivation of Bethe's
result for a small isolated hole in a metallic screen \cite{B1944}
follows from this principle: for incidence normal to a metallic
disk of diameter $D$, the electric polarizability reduces to
$\alpha_E=D^3/6\pi$, while the magnetic response vanishes
altogether \cite{J1975}, leading to a forward scattering cross
section given by $\sigma/(\pi D^2/4)=4 (kD)^4/27\pi^2$ (normalized
to the disk area), as calculated from the Poynting vector in the
forward hemisphere, where $k=2\pi/\lambda$; now, Babinet's
principle implies that this result must coincide with the
transmission cross section of a hole of the same diameter, in
perfect agreement with Bethe's formula \cite{B1944}.



For hole arrays in PCTFs, Babinet's principle connects their
transmittance to the reflectance of the complementary geometries
consisting of planar arrays of perfect-conductor thin disks
\cite{Eggimann}. In the small-hole limit, this can be obtained by
self-consistently solving for the electric dipoles induced on the
disks, ${\bf p}_{\bf R}=\alpha_E {\bf E}^{\rm inc}({\bf
R})+\alpha_E \sum_{{\bf R}'\neq {\bf R}} G_{E}^{{\bf RR}'} {\bf
p}_{{\bf R}'}$, where ${\bf R}$ and ${\bf R}'$ label disk sites,
$\alpha_E$ is the electric polarizability of the disks, and
$G_{E}^{{\bf RR}'}$ is the dipole-dipole interaction dyadic. At
normal incidence on periodic arrays, the polarization is the same
in all disks, parallel to the incident electric field ${\bf
E}^{\rm inc}$ ($\parallel \hat{\bf x}$), and given by
\begin{eqnarray}
p=\frac{1}{\frac{1}{\alpha_E}-G_E}, \label{ppp}
\end{eqnarray}
where
\begin{eqnarray} G_E=\sum_{{\bf R}\neq 0}(k^2+\partial_{xx}^2){\rm e}^{{\rm i}kR}/R. \label{GE}
\end{eqnarray}
Below the threshold of the first diffraction channel, the
far-field induced by the lattice of identical dipoles is readily
calculated as $E^{\rm ind}=(2\pi{\rm i}k/A) p \;{\rm e}^{{\rm
i}k|z|}$, where $A$ is the area of the unit cell and $z$ is chosen
perpendicular to the lattice plane. From here, one obtains the
transmittance $|1+(2\pi{\rm i}k/A) p|^2$ and reflectance $|(2\pi
k/A) p|^2$ of the disk array, which have to sum 1 because perfect
conductor cannot dissipate energy. This condition leads to ${\rm
Im}\{1/\alpha_E - G_{E}\}=-2 \pi k/A$, which generalizes the
optical theorem for non-absorbing particles, ${\rm
Im}\{-1/\alpha_E\}=2 k^3/3$. We can then write the reflectance of
the disk array, which equals the transmittance of the
complementary hole array from Babinet's principle, as
\begin{eqnarray}
T=\frac{1}{1+(\frac{A}{2\pi k}{\rm Re}\{\alpha_E^{-1}-G_E\})^2}.
\label{dipoles}
\end{eqnarray}
Like in the Wood anomalies \cite{W1935,SVV03}, the structure
factor $G_E$ diverges when one of the reflected beams goes
grazing, and in particular near $\lambda=a$, the period for square
lattices. More precisely, ${\rm Re}\{G_E
a^3\}=4\pi^2\sqrt{2}/\sqrt{\lambda/a-1} + C(\lambda/a)$, with
$\lambda>a$, where the first term is derived analytically from the
divergent terms of the sum over reciprocal lattice vectors into
which Eq.\ (\ref{GE}) can be recast, and $C(\theta) \approx 35
\,\exp[-22(\theta-1)]-118$ is a smooth fit to the remaining
non-divergent terms.

\begin{figure}
\centerline{\scalebox{0.45}{\includegraphics{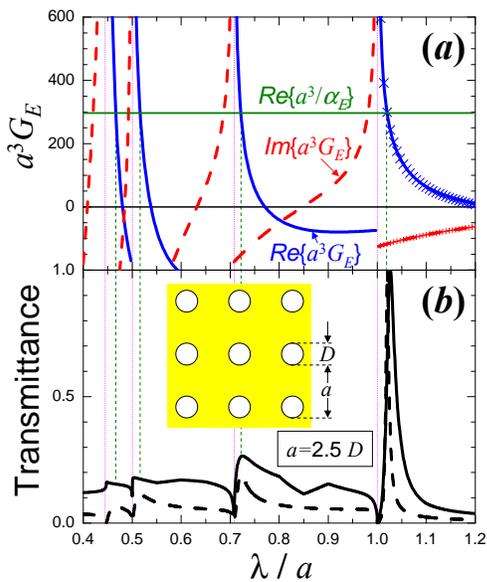}}}
\caption{\label{Fig2} {\bf (a)} Wavelength dependence of $G_E$ [Eq.\
(\ref{GE})] and determination of transmission maxima and minima: the
solid (broken) curve represents the real (imaginary) part of $G_E$
for a square lattice under normal incidence. Transmission minima
(see vertical dotted lines) result from divergences of $G_E$, while
transmission maxima (vertical dashed lines) are derived from the
condition that ${\rm Re}\{G_E\}$ equals the inverse of the
polarizability of the hole. {\bf (b)} Transmittance under the same
conditions as in Fig.\ \ref{Fig1}(a) over a wider wavelength range.}
\end{figure}

Obviously, $G_E$ diverges as $\lambda\rightarrow a^+$, and thus,
given an arbitrarily-small hole diameter, there is always one
wavelength $\lambda>a$ for which the second term in the
denominator of Eq.\ (\ref{dipoles}) vanishes, implying that the
transmittance (reflectance) becomes 100\% at that frequency for
the array of holes (disks), as shown in Fig.\ \ref{Fig1} (dashed
curves). This is graphically illustrated in Fig.\ \ref{Fig2},
where ${\rm Re}\{G_E\}$ has been represented [Fig.\ \ref{Fig2}(a)]
and compared with ${\rm Re}\{1/\alpha_E\}$ (horizontal line). The
crossing points of these two agree very well with the transmission
maxima predicted by the analytical Eq.\ (\ref{dipoles}) [dashed
curve in Fig.\ \ref{Fig2}(b)]. Our findings support recent
experiments performed at THz frequencies, where nearly
perfect-conductor behavior is expected,
and up to 90\% absolute transmission has been observed
\cite{MH04}.

This divergence of $G_E$ is the same one that produces Wood's
anomalies and is related to the geometry of the system. For disks,
this is simply coming from the accumulation of in-phase
contributions from long-distance coherent multiple scattering (MS)
\cite{T99,SVV03,LT04}. In the complementary structure (hole
arrays), the same systematics is at work via Babinet's principle,
as described above.
Moreover, the reflectance (transmittance) of the disk (hole) array
[Eq.\ (\ref{dipoles})] exhibits the same behavior as Fano
reflection resonances,
in agreement with reported descriptions of the
mechanism of transmission enhancement \cite{SVV03}. In general,
Fano resonances \cite{F1961} occur when the energy (frequency) of
the incoming wave is tuned to the energy (frequency) of a QBS. Our
QBS is confined to the lattice and partially leaking into the
continuum of vacuum light states, giving rise to some broadening
of the transmission peaks, as observed in Fig.\ \ref{Fig1}.


For perfect-conductor disks or hole arrays, we can then consider
the transmission enhancement as a consequence of the coupling of
the incoming field with a QBS of {\em geometrical} origin. Whether
this is or is not denoted surface-plasmon constitutes in our
opinion a matter of taste regarding terminology.





We have also carried out a rigorous solution of Maxwell's
equations that is represented in Fig.\ \ref{Fig1} by solid curves.
In our method, we first obtain the multipolar scattering matrix of
a disk of the same diameter as our holes using the boundary
element method \cite{paper030}. Then, we solve the problem of a
lattice of disks, which is the exact complementary screen geometry
of our hole array, using the layer-KKR approach \cite{SYM98}.
Finally, the solution for the hole array is related to the
solution for the disk array using Babinet's principle, and in
particular, the transmittance for s (p) polarization in the former
equals the reflectance for p (s) polarization in the latter.


Fig.\ \ref{Fig1} shows results for two different ratios of the
diameter $D$ to the lattice constant $a$. The analytical model of
Eq.\ (\ref{dipoles}) explains well the shift of the transmission
resonance wavelength towards $a$ and its narrowing with decreasing
$D/a$. Actually, Eq.\ (\ref{dipoles}) works better for smaller
holes, since it is based upon the dipolar part of their response,
and it should be exact in the $D/a\rightarrow 0$ limit, although
higher perfectness of hole shape, size, and position will be
necessary to observe narrower resonances.

Of course, the derivation of Eq.\ (\ref{dipoles}) fails when more
than one transmitted beam is present: 100\% transmission can only
occur for $\lambda>a$ in our square lattices. This is shown in the
transmission maxima of Fig.\ \ref{Fig2}(b) for $\lambda<a$, where
non-normal beams take part of the light flux, and where Eq.\
(\ref{dipoles}) (broken curve) is shown to reproduce qualitatively
our rigorous calculations (solid curve).

\begin{figure}
\centerline{\scalebox{0.30}{\includegraphics{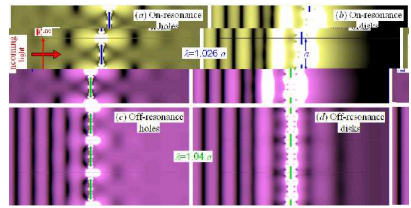}}}
\caption{\label{Fig3} {\bf (a)} and {\bf (c)}: squared electric
field in log scale for light incident normal to the hole array
considered in Fig.\ \ref{Fig1}(a), under resonance and non-resonance
conditions, respectively (see arrows in [Fig.\ \ref{Fig1}(a)]. The
plane of representation is parallel both to the screen normal and to
one of the lattice vectors, and it intersects the centers of a row
of holes. The metallic screen and its holes are represented by
vertical broken lines. {\bf (b)} and {\bf (d)}: same for the disk
array defined as the complementary screen with respect to that of
(a) and (c).}
\end{figure}

A closer look into the near field is given in Fig.\ \ref{Fig3} for
on-resonance [(a) and (b)] and off-resonance [(c) and (d)]
scenarios in a hole array [(a) and (c)] and in its complementary
disk array [(b) and (d)]. The effect of the on-resonance hole
array [Fig.\ \ref{Fig3}(a)] extends up to a distance of the order
of the lattice period $a$, beyond which the electric field
strength is quite uniform. For the on-resonance disk array [Fig.\
\ref{Fig3}(b)], there is an evanescent field beyond the plane of
the disks and total reflection establishes an interference pattern
on the incoming-light side. The off-resonance scenario exhibits
interference of incoming and partly-reflected light, as well as
partial light transmission. It should be noticed that the field
near the hole array is more intense on resonance (this is a
signature of a QBS) as a result of stronger MS, and this
anticipates stronger absorption if lossy materials are employed
rather than perfect conductors, in agreement with recent
experiments \cite{BMD04}.

When light is coming under oblique incidence, the above model has
to be supplemented by adding a phase $\exp\{{\rm i}{\bf k}^{\rm
inc}_\parallel\cdot{\bf R}\}$ to the dipoles induced on the
different hole sites ${\bf R}$. And more importantly, both
magnetic and electric responses will play a role. In fact, any
perfect-conductor planar object of finite extension such as our
disks can sustain induced currents and charges only within the
plane where it is contained, leading to induced electric dipoles
parallel to the plane and induced magnetic dipoles perpendicular
to it in the long wavelength limit. The latter can only be excited
under oblique incidence.

\begin{figure}
\centerline{\scalebox{0.55}{\includegraphics{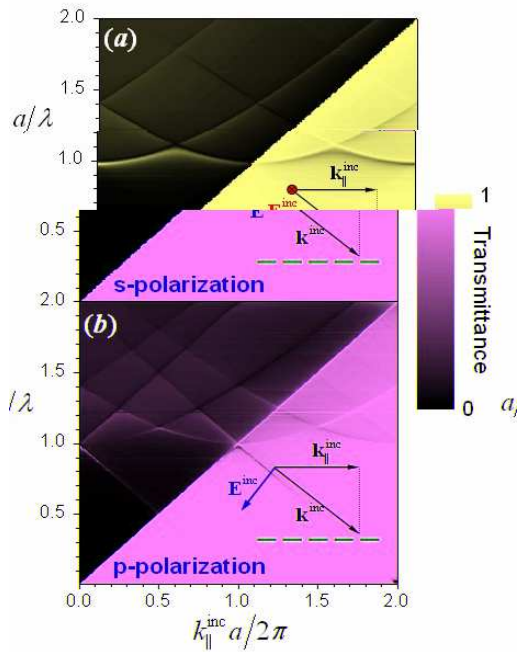}}}
\caption{\label{Fig4} Angular dependence of the transmission of s-
and p-polarized light [region above the light line in {\bf (a)} and
{\bf (b)}, respectively] incident upon a perfect-conductor screen
perforated by an array of circular holes of diameter $D=a/2.5$
arranged in a square lattice of constant $a$. Resonances outside the
light cone are also shown by plotting the transmitted far field when
the screen is exposed to an incident evanescent wave.}
\end{figure}

In particular, the magnetic polarizability of a disk,
$\alpha_M=-D^3/12\pi$, reacts only to normal magnetic field
components. The disks in a lattice are coupled via $G_M=\sum_{{\bf
R}\neq 0}\exp\{{\rm i}{\bf k}^{\rm inc}_\parallel\cdot{\bf
R}\}(k^2+\partial_{zz}^2){\rm e}^{{\rm i}kR}/R$ to produce
self-consistent perpendicular magnetic dipoles given by the
magnetic counterpart of Eq.\ (\ref{ppp}). Therefore, there are two
uncoupled sets of resonances in a disk array that respond to
external (i) perpendicular magnetic fields and (ii) parallel
electric fields, which in virtue of Babinet's principle are
translated into hole-array resonances excited by external (i)
perpendicular electric fields and (ii) parallel magnetic fields,
respectively. For s-polarized incident light (${\bf E}^{\rm inc}$
parallel to the screen) only the latter can be excited, whereas
p-polarized light can couple to both types of resonances,
as illustrated in Fig.\ \ref{Fig4} via the dependence of the
transmittance on incident-light wavelength and parallel momentum.
The region above the light line ($\lambda<k^{\rm inc}_\parallel$)
shows transmission maxima (bright areas) and minima (right above
the maxima), with 100\% transmission achieved only for p
polarization and $\lambda>a$. As expected, p-polarized light
couples to more resonances [of types (i) and (ii)] than
s-polarized light [only resonances of type (ii)]. This conclusions
might be relevant to explain recent measurements of polarization-
and angle-resolved transmission in thick arrays of real metal
\cite{BMD04} with similar qualitative behavior as that of Fig.\
\ref{Fig4}.

Our methods can be trivially extended to other geometries like
finite arrays \cite{MH04} and more complex hole shapes
\cite{KES04,GBM04}. In particular, the asymmetry of the
polarizability tensor of a rectangular metallic plate, which is
larger along its long axis, is consistent with the stronger red
shift of the transmitted maxima observed for lattices of square
holes \cite{KES04} as compared to circular holes of similar size.
Also, new metamaterials composed of layers of hole arrays and disk
arrays could be analyzed by straightforward generalization of the
methods presented above.

In real metals, the dispersion relation of surface plasmons
deviates from the light line and it is plausible that the
transmission enhancement observed in the NIR is the result of
in-phase MS mediated by electromagnetic propagation via these
modes.

Furthermore, the metal films are supported on glass substrates in
many experiments \cite{ELG98,KES04,BMD04}. This analysis can be
easily extended to a PCTF surrounded by dielectrics of different
index of refraction on either side as shown by Dawes {\it et al.}
\cite{Dawes}, in which case full transmission resonances occur
only when the wavelength in the higher-index region is above the
smallest reciprocal surface lattice vector of the array (e. g.,
near $\lambda\approx 1.4 a$ for a square lattice supported on
glass), and the effective {\it polarizability} of the hole differs
from the symmetric case considered above, but apart from this the
long-wavelength analysis presented above remains the same.


In summary, we have shown that the extraordinary transmission of
light through hole arrays in PCTFs can be fully explained in
simple terms by invoking Babinet's principle. The existence of
100\% transmission maxima has been established, regardless how
small the holes are. The origin of full transmission can be found
in phase accumulation during long-range dipole-dipole interaction
among holes when the wavelength is close to the onset of of the
first diffraction order. Potential application to narrow filters
can be envisioned in the microwave and THz regimes, since the
transmission resonances are increasingly shaper as the holes
become smaller.




\begin{thebibliography}{24}
\expandafter\ifx\csname
natexlab\endcsname\relax\def\natexlab#1{#1}\fi
\expandafter\ifx\csname bibnamefont\endcsname\relax
  \def\bibnamefont#1{#1}\fi
\expandafter\ifx\csname bibfnamefont\endcsname\relax
  \def\bibfnamefont#1{#1}\fi
\expandafter\ifx\csname citenamefont\endcsname\relax
  \def\citenamefont#1{#1}\fi
\expandafter\ifx\csname url\endcsname\relax
  \def\url#1{\texttt{#1}}\fi
\expandafter\ifx\csname
urlprefix\endcsname\relax\def\urlprefix{URL }\fi
\providecommand{\bibinfo}[2]{#2}
\providecommand{\eprint}[2][]{\url{#2}}

\bibitem[{\citenamefont{Ebbesen et~al.}(1998)\citenamefont{Ebbesen, Lezec,
  Ghaemi, Thio, and Wolff}}]{ELG98}
\bibinfo{author}{\bibfnamefont{T.~W.} \bibnamefont{Ebbesen}} \bibnamefont{{\em et al.}},
  \bibinfo{journal}{Nature} \textbf{\bibinfo{volume}{391}},
  \bibinfo{pages}{667} (\bibinfo{year}{1998});
\bibinfo{author}{\bibfnamefont{H.~F.} \bibnamefont{Ghaemi}} \bibnamefont{{\em et al.}},
  \bibinfo{journal}{Phys.\ Rev.\ B} \textbf{\bibinfo{volume}{58}},
  \bibinfo{pages}{6779} (\bibinfo{year}{1998}).

\bibitem[{\citenamefont{Bethe}(1944)}]{B1944}
\bibinfo{author}{\bibfnamefont{H.~A.} \bibnamefont{Bethe}},
  \bibinfo{journal}{Phys.\ Rev.} \textbf{\bibinfo{volume}{66}},
  \bibinfo{pages}{163} (\bibinfo{year}{1944}).

\bibitem[{\citenamefont{{G\'{o}mez-Rivas}
  et~al.}(2003)\citenamefont{{G\'{o}mez-Rivas}, Schotsch, {Haring Bolivar}, and
  Kurz}}]{GSH03}
\bibinfo{author}{\bibfnamefont{J.}~\bibnamefont{{G\'{o}mez-Rivas}}} \bibnamefont{{\em et al.}},
  \bibinfo{journal}{Phys.\ Rev.\ B} \textbf{\bibinfo{volume}{68}},
  \bibinfo{pages}{201306(R)} (\bibinfo{year}{2003}).

\bibitem[{\citenamefont{{Klein Koerkamp} et~al.}(2004)\citenamefont{{Klein
  Koerkamp}, Enoch, Segerink, {van Hulst}, and Kuipers}}]{KES04}
\bibinfo{author}{\bibfnamefont{K.~J.} \bibnamefont{{Klein Koerkamp}}} \bibnamefont{{\em et al.}},
  \bibinfo{journal}{Phys.\ Rev.\ Lett.} \textbf{\bibinfo{volume}{92}},
  \bibinfo{pages}{183901} (\bibinfo{year}{2004}).

\bibitem[{\citenamefont{Lezec and Thio}(2004)}]{LT04}
\bibinfo{author}{\bibfnamefont{H.~J.} \bibnamefont{Lezec}} \bibnamefont{and}
  \bibinfo{author}{\bibfnamefont{T.}~\bibnamefont{Thio}},
  \bibinfo{journal}{Opt.\ Express} \textbf{\bibinfo{volume}{12}},
  \bibinfo{pages}{3629} (\bibinfo{year}{2004}).

\bibitem[{\citenamefont{Gordon et~al.}(2004)\citenamefont{Gordon, Brolo,
  McKinnon, Rajora, Leathem, and Kavanagh}}]{GBM04}
\bibinfo{author}{\bibfnamefont{R.}~\bibnamefont{Gordon}} \bibnamefont{{\em et al.}},
  \bibinfo{journal}{Phys.\ Rev.\ Lett.} \textbf{\bibinfo{volume}{92}},
  \bibinfo{pages}{037401} (\bibinfo{year}{2004}).

\bibitem[{\citenamefont{Barnes et~al.}(2004)\citenamefont{Barnes, Murray,
  Dintinger, Devaux, and Ebbesen}}]{BMD04}
\bibinfo{author}{\bibfnamefont{W.~L.} \bibnamefont{Barnes}} \bibnamefont{{\em et al.}},
  \bibinfo{journal}{Phys.\ Rev.\ Lett.} \textbf{\bibinfo{volume}{92}},
  \bibinfo{pages}{107401} (\bibinfo{year}{2004}).

\bibitem[{\citenamefont{Miyamaru and Hangyo}(2004)}]{MH04}
\bibinfo{author}{\bibfnamefont{F.}~\bibnamefont{Miyamaru}} \bibnamefont{and}
  \bibinfo{author}{\bibfnamefont{M.}~\bibnamefont{Hangyo}},
  \bibinfo{journal}{Appl.\ Phys.\ Lett.} \textbf{\bibinfo{volume}{84}},
  \bibinfo{pages}{2742} (\bibinfo{year}{2004}).

\bibitem[{\citenamefont{Cao and Nahata}(2004)}]{CN04}
\bibinfo{author}{\bibfnamefont{H.}~\bibnamefont{Cao}} \bibnamefont{and}
  \bibinfo{author}{\bibfnamefont{A.}~\bibnamefont{Nahata}},
  \bibinfo{journal}{Opt.\ Express} \textbf{\bibinfo{volume}{12}},
  \bibinfo{pages}{1004} (\bibinfo{year}{2004}).

\bibitem[{\citenamefont{Treacy}(1999)}]{T99}
\bibinfo{author}{\bibfnamefont{M.~M.~J.} \bibnamefont{Treacy}},
  \bibinfo{journal}{Appl.\ Phys.\ Lett.} \textbf{\bibinfo{volume}{75}},
  \bibinfo{pages}{606} (\bibinfo{year}{1999});
  \bibinfo{journal}{Phys.\ Rev.\ B} \textbf{\bibinfo{volume}{66}},
  \bibinfo{pages}{195105} (\bibinfo{year}{2002}).

\bibitem[{\citenamefont{Popov et~al.}(2000)\citenamefont{Popov, Nevi\`{e}re,
  Enoch, and Reinisch}}]{PNE00}
\bibinfo{author}{\bibfnamefont{E.}~\bibnamefont{Popov}} \bibnamefont{{\em et al.}},
  \bibinfo{journal}{Phys.\ Rev.\ B} \textbf{\bibinfo{volume}{62}},
  \bibinfo{pages}{16100} (\bibinfo{year}{2000}).

\bibitem[{\citenamefont{{Mart\i{\i}n-Moreno}
  et~al.}(2001)\citenamefont{{Mart\'{\i}n-Moreno}, {Garc\'{\i}a-Vidal}, Lezec,
  Pellerin, Thio, Pendry, and Ebbesen}}]{MGL01}
\bibinfo{author}{\bibfnamefont{L.}~\bibnamefont{{Mart\'{\i}n-Moreno}}} \bibnamefont{{\em et al.}},
  \bibinfo{journal}{Phys.\ Rev.\ Lett.}
  \textbf{\bibinfo{volume}{86}}, \bibinfo{pages}{1114} (\bibinfo{year}{2001}).

\bibitem[{\citenamefont{Wannemacher}(2001)}]{W01}
\bibinfo{author}{\bibfnamefont{R.}~\bibnamefont{Wannemacher}},
  \bibinfo{journal}{Opt. Commun.} \textbf{\bibinfo{volume}{195}},
  \bibinfo{pages}{107} (\bibinfo{year}{2001}).

\bibitem[{\citenamefont{Salomon et~al.}(2001)\citenamefont{Salomon, Grillor,
  Zayats, and {de Fornel}}}]{SGZ01}
\bibinfo{author}{\bibfnamefont{L.}~\bibnamefont{Salomon}} \bibnamefont{{\em et al.}},
  \bibinfo{journal}{Phys.\ Rev.\ Lett.}
  \textbf{\bibinfo{volume}{86}}, \bibinfo{pages}{1110} (\bibinfo{year}{2001}).

\bibitem[{\citenamefont{Sarrazin et~al.}(2003)\citenamefont{Sarrazin, Vigneron,
  and Vigoureux}}]{SVV03}
\bibinfo{author}{\bibfnamefont{M.}~\bibnamefont{Sarrazin}},
  \bibinfo{author}{\bibfnamefont{J.-P.} \bibnamefont{Vigneron}},
  \bibnamefont{and} \bibinfo{author}{\bibfnamefont{J.-M.}
  \bibnamefont{Vigoureux}}, \bibinfo{journal}{Phys.\ Rev.\ B}
  \textbf{\bibinfo{volume}{67}}, \bibinfo{pages}{085415}
  (\bibinfo{year}{2003}).


\bibitem[{\citenamefont{Jackson}(1975)}]{J1975}
\bibinfo{author}{\bibfnamefont{J.~D.} \bibnamefont{Jackson}},
  \emph{\bibinfo{title}{Classical Electrodynamics}}
  (\bibinfo{publisher}{Wiley}, \bibinfo{address}{New York},
  \bibinfo{year}{1975}).

\bibitem[{\citenamefont{Wood}(1935)}]{W1935}
\bibinfo{author}{\bibfnamefont{R.~W.} \bibnamefont{Wood}},
  \bibinfo{journal}{Phys.\ Rev.} \textbf{\bibinfo{volume}{48}},
  \bibinfo{pages}{928} (\bibinfo{year}{1935}).

\bibitem[{\citenamefont{Fano}(1961)}]{F1961}
\bibinfo{author}{\bibfnamefont{U.}~\bibnamefont{Fano}},
  \bibinfo{journal}{Phys.\ Rev.} \textbf{\bibinfo{volume}{124}},
  \bibinfo{pages}{1866} (\bibinfo{year}{1961}).

\bibitem[{Eggimann()}]{Eggimann}
\bibinfo{note}{W.~H. Eggimann and R.~E. Collin, IRE Trans.
Micr. Theory and Tech. {\bf 10}, 528 (1962); C.~C. Chen, IEEE
Trans. Micr. Theory and Tech. {\bf 19}, 475 (1971).}

\bibitem[{McPhedran()}]{McPhedran}
\bibinfo{note}{R.~C. McPhedran {\it et al.},
in {\it Electromagnetic Theory of Gratings}, Ed. by R. Petit, p.
227 (Springer, Berlin, 1980).}





\bibitem[{\citenamefont{{Garc\'{\i}a de Abajo} and Howie}(1998)}]{paper030}
\bibinfo{author}{\bibfnamefont{F.~J.} \bibnamefont{{Garc\'{\i}a de Abajo}}}
  \bibnamefont{and} \bibinfo{author}{\bibfnamefont{A.}~\bibnamefont{Howie}},
  \bibinfo{journal}{Phys.\ Rev.\ Lett.} \textbf{\bibinfo{volume}{80}},
  \bibinfo{pages}{5180} (\bibinfo{year}{1998}).


\bibitem[{SYM()}]{SYM98}
\bibinfo{note}{N. Stefanou, V. Yannopapas, and A. Modinos, Comput. Phys.
  Commun. {\bf 113}, 49 (1998); {\bf 132}, 189 (2000).}

\bibitem[{Dawes()}]{Dawes}
\bibinfo{note}{D.~H. Dawes, R.~C. McPhedran, and L.~B. Whitbourn, Appl. Opt.
{\bf 28}, 3498 (1989 ).}

\end{thebibliography}
\end{document}